\documentclass[fleqn,twoside]{article}
\usepackage{espcrc2}

\usepackage{epsfig}

\newif\ifpreprint
\preprinttrue

\title{
\ifpreprint
\rightline{\normalsize FERMILAB-Conf-02/181-E}
\rightline{\normalsize UASLP-IF-02-007}
\vspace{0.2cm}
SELEX RICH Performance and Physics Results\thanks{Contributed talk at the
Fourth Workshop on
RICH Detectors, June 5-10, 2002, Pylos, Greece. To be published in NIMA.}
\else
SELEX RICH Performance and Physics Results
\fi
}

\author{J.~Engelfried\address[fermi]{Fermi National Accelerator Laboratory,
Batavia, IL, USA}\address[slp]{Instituto de F\'{\i}sica, 
Universidad Aut\'onoma de San Luis Potos\'{\i}, Mexico}\thanks{Corresponding author; email:{\tt jurgen@ifisica.uaslp.mx}},
I.S.~Filimonov\address[msu]{Moscow State University,
Moscow, Russia}\thanks{deceased},
J.~Kilmer\addressmark[fermi], 
A.P.~Kozhevnikov\address[ihep]{Institute for High Energy Physics,
Protvino, Russia},
V.P.~Kubarovsky\addressmark[ihep], 
V.V.~Molchanov\addressmark[ihep],
A.V~Nemitkin\addressmark[msu], 
E.~Ramberg\addressmark[fermi], 
V.I.~Rud\addressmark[msu],
L.~Stutte\addressmark[fermi]
}

\begin{document}

\begin{abstract}
SELEX took data in the 1996/7 Fixed Target Run at Fermilab.  The excellent
performance parameters of the SELEX RICH Detector had direct
influence on the quality of the obtained physics results.
\end{abstract}

\maketitle

\section{Introduction}
The Fermilab experiment E781 (SELEX): 
A Segmented Large $x_F$ Baryon Spectrometer~\cite{e781,SELEX},
which took 
data in the 1996/97 fixed target run at Fermilab, is designed to
perform high statistics studies of production mechanisms and decay physics of
charmed and charmed-strange baryons such as
$\Sigma_c$, $\Xi_c$, $\Omega_c$ and $\Lambda_c$\@.
The physics goals of the experiment require good charged
particle identification to look for the different baryon decay modes.
One must be able to separate $\pi$, $K$ and $p$
over a wide momentum range when looking for charmed baryon decays like
$\Lambda_c^+ \rightarrow p \, K^- \pi^+$.

A RICH detector with a 2848~phototube photocathode array
has been constructed~\cite{proto,sphinx} to do this. The detector begins
about $16\,\mbox{m}$ downstream of the charm production target, with two
analysis magnets with $800\,\mbox{MeV/c}$ $p_t$-kick each in-between, and is
surrounded by multi-wire proportional and drift chambers which provide particle
tracking.  The average number of tracks reaching the RICH is about 5 per event.
First results from this detector can be found in~\cite{elba}. A detailed
description of the detector itself is given in~\cite{bignim,israel}.

In this article we first describe shortly the main 
parts of the detector
(vessel, mirrors, photon detector), after this we will review
its stability during the run and performance for physics analyses.
 Finally we report about physics results obtained with the help of the RICH
up to now.

\section{Description of the detector}
A detailed description can be found in~\cite{bignim}. Here we will
repeat the main features only.
The SELEX RICH detector consists of a $10.22\,\mbox{m}$ long vessel
with $93\,\mbox{in.\ }$diameter, filled with pure Neon at atmospheric
pressure. After initial filling to a slight ($1\,\mbox{psi}$) overpressure, the
vessel was perfectly tight for a period of more than 15 months.

At the end of the vessel an array of 16 hexagonal spherical mirrors is mounted 
on a low-mass hex-panel with 3-point mounts to form a sphere of
$19.8\,\mbox{m}$ radius.  Each mirror is $10\,\mbox{mm}$ thick,
made out of low-expansion class, 
and has a reflectivity $>85\,\mbox{\%}$ at $160\,\mbox{nm}$.
Before evaporating, the quality of every mirror was measured~\cite{Ronchi}.

The phototube matrix at the focal plane consists of 2848 
$1/2\,\mbox{in.\ }$photomultipliers in a hexagonal closed packed matrix of
89~columns in 32~rows.
One column of 32 phototubes is connected to one common high voltage.
In the center part, alternating columns of Hamamatsu R760 and FEU60 tubes
are employed, and in the outer part only FEU60 tubes are used.
Every phototube is connected to a preamplifier-discriminator-ECL driver hybrid
chip, and the readout of the signals is performed with a standard wire-chamber
electronics~\cite{CROS} with $170\,\mbox{nsec}$ integration time.

The whole vessel is tilted by $2.4^\circ$ so particles are not passing
through the phototubes.

\section{Performance of detector}
The detector performance over the 15 month running period was very stable.

In the central region (mixed R760 and FEU60) we observed on 
average 13.6~photons for a $\beta=1$
particle, corresponding to a figure of merit~\cite{desrich} 
$N_0=104\,\mbox{cm}^{-1}$, and in the 
outside (only FEU60) we obtained $N_0=70\,\mbox{cm}^{-1}$.

The refractive index, measured in several different ways, was stable
to better than $0.6\,\mbox{\%}$,
which would change the ring radius at $\beta=1$ by
less than $10\,\mbox{\%}$ of its resolution.

The single photon resolution was estimated to be $5.5\,\mbox{mm}$, with the
biggest contribution coming from the size of the phototubes.  The ring
radius resolution for the typical multitrack environment was measured at
$1.56\,\mbox{mm}$.
\begin{figure}
\leavevmode
\begin{center}
\epsfxsize=\hsize
\ifpreprint
\epsffile{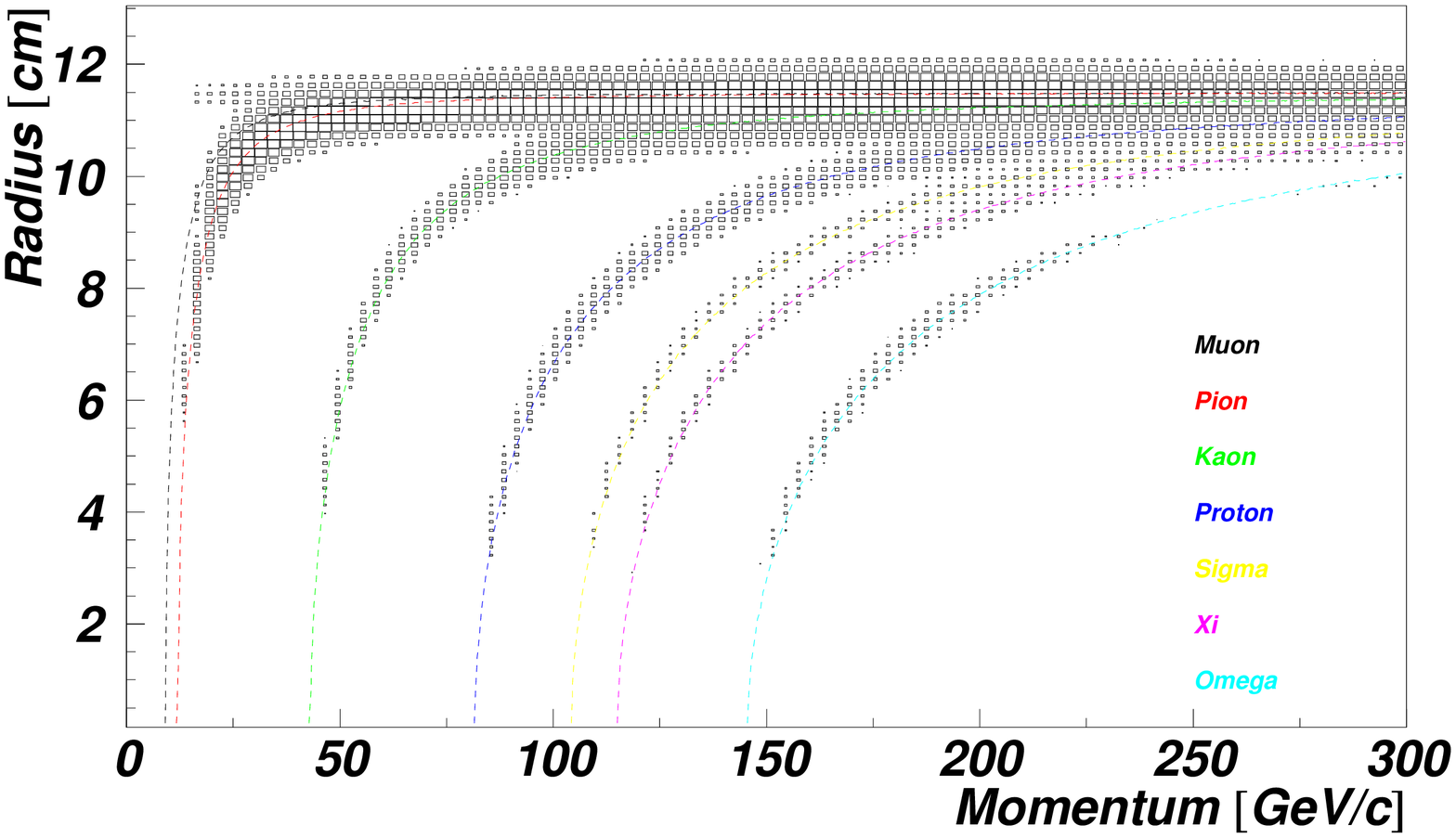}
\else
\epsffile{radiusa.eps}
\fi
\epsfxsize=\hsize
\ifpreprint
\epsffile{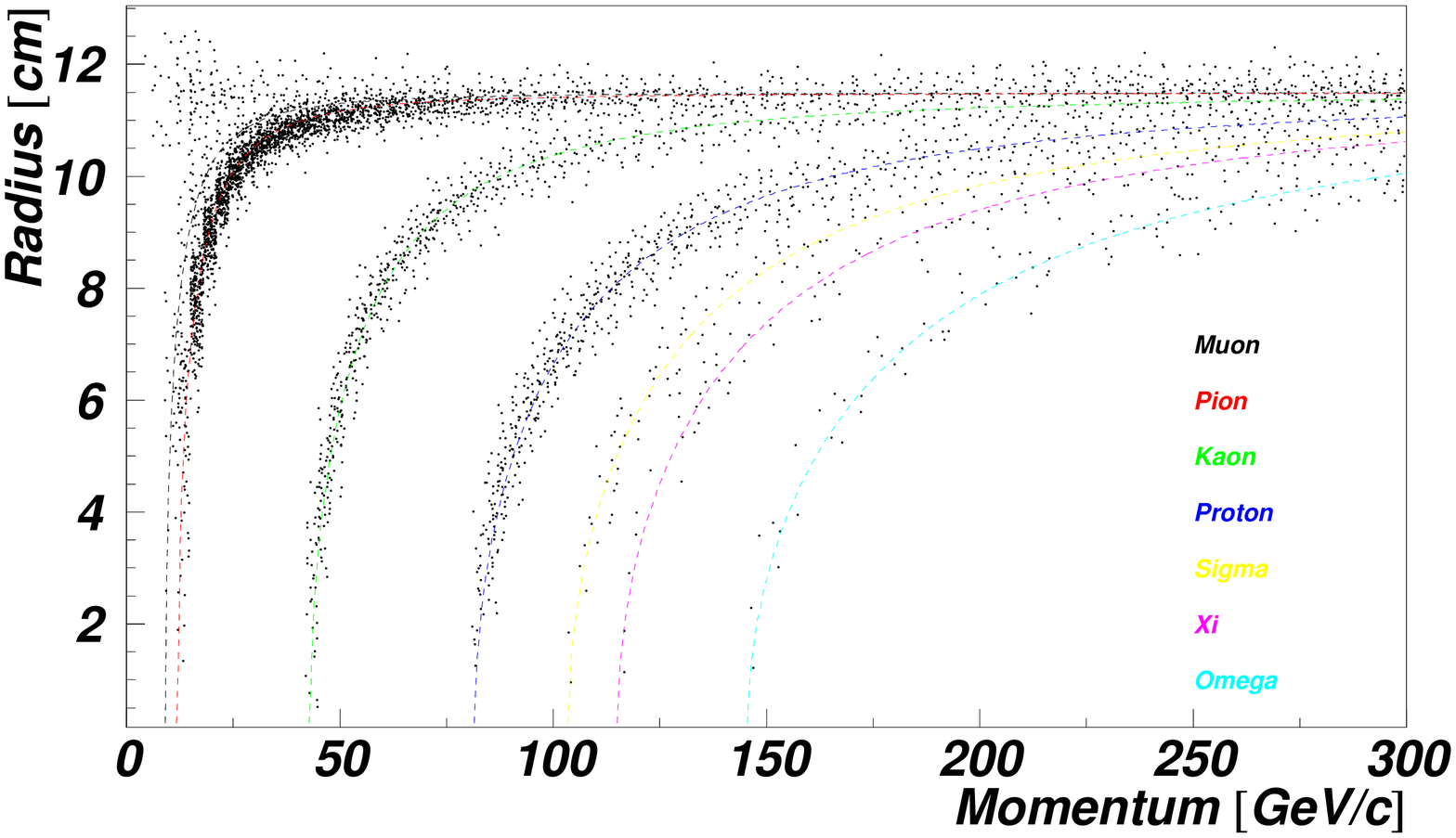}
\else
\epsffile{radiusb.eps}
\fi
\caption{Distribution of ring radii and momentum for single track
events. Top: $53\cdot10^6$ negative tracks. 
Bottom: 180000 positive tracks.
The lines show the expected for ring radii for $\mu^\pm$, $\pi^\pm$,
$K^\pm$, $p^\pm$, $\Sigma^\pm$, $\Xi^\pm$, and $\Omega^\pm$.}
\label{radii}
\end{center}
\end{figure}
In fig.~\ref{radii} we show the distribution of ring radii for single
track events, where one can identify eight particles and eight anti-particles,
demonstrating the huge dynamic range of this detector.
With the same dataset we measured the noise of the detector, by counting
the number of hits not assigned to a ring, taking into account the
integration time of the readout electronics.
\begin{figure}
\leavevmode
\begin{center}
\epsfxsize=\hsize
\epsffile{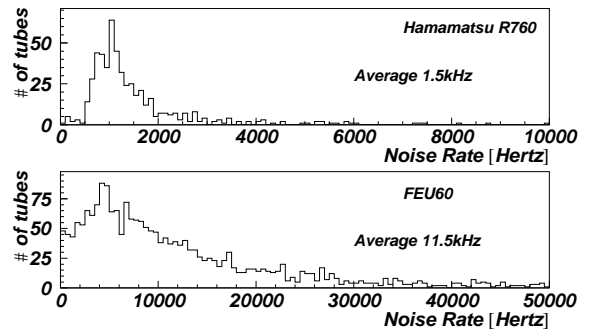}
\caption{Distribution of noise rates (conservative estimate)
for the two types of phototubes used in the detector.} 
\label{noise}
\end{center}
\end{figure}
The result is shown in fig.~\ref{noise}, for the two different types
of phototubes used.  For beam-off events we observed on average 6~hits. 

For particle identification, we calculate likelihoods for different
particle hypotheses~\cite{likeli}, using information from the tracking
system.
With the help of kinematically reconstructed signals we measure the
identification efficiency for different cuts. For protons
(using $\Lambda\to p\pi^-$ decays) in the
momentum range $100\,\mbox{GeV}/c<p_p<200\,\mbox{GeV}/c$, the
identification efficiency is around $98\,\mbox{\%}$, and even below 
proton threshold ($\approx90\,\mbox{GeV}/c$), the efficiency is above
$90\,\mbox{\%}$~\cite{bignim},
with a mis-identification rate of a few percent, which can
be attributed to tracking errors.  The same is true for kaons~\cite{bignim}.

\begin{figure}
\leavevmode
\begin{center}
\epsfxsize=\hsize
\epsffile{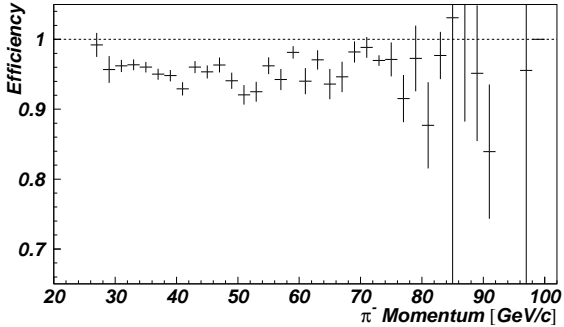}
\caption{Identification efficiency for $\pi^-$ as function of momentum.}
\label{pions}
\end{center}
\end{figure}
Pions from charm decays have usually lower momenta than other decay products;
for that reason in SELEX we accept as pions in charm analysis also
tracks which do not reach the RICH.  Should the track be within the RICH
acceptance, we require the ratio of the likelihood of that track to be
a pion to all other particles to be $\ge0.1$.
In fig.~\ref{pions} we show that the identification efficiency is above
$90\,\mbox{\%}$ without a sharp turn-on.

\section{Physics Results}
SELEX submitted first publications in 1999, after first reports at
conferences in 1998.  To date SELEX has published seven papers and two more 
are submitted.  Some papers on mesons and hyperons do not use the RICH
in their analyses~\cite{uwe,a2,sigmaradius,aharon}, but for charm physics
the RICH proved to be essential in the analysis, reducing the background
significantly but still having high efficiency.

SELEX published measurements of the $\Lambda_c^+$~\cite{lclife}
and $D_s^\pm$~\cite{dslife} lifetimes, where the RICH was essential 
in suppressing reflection below the mass peak stemming from mis-identified
daughter particles.

The measurement of production properties
of $\Lambda_c^\pm$~\cite{lcprod}
 with 4 different beam particles ($\Sigma^-$, $\pi^-$, 
$p$, $\pi^+$) also benefitted largely from the RICH.
More hadroproduction results for $D^0$, $D^\pm$, and $D_s^\pm$ are
in preparation~\cite{Zacatecas}.

SELEX also observed for the first time a Cabbibo suppressed decay mode
of a charmed strange baryon, the $\Xi_c^+\to pK^-\pi^+$~\cite{cabsupxi}.

Just recently, SELEX reported on the first observation of doubly-charmed
baryons~\cite{ccu}.

\section{Summary and Outlook}
The RICH detector is an essential part of the physics analysis in SELEX,
due to its high efficiency and large dynamic range.  It fulfilled all
the design criteria.

With seven papers already published plus two submitted, 
the collaboration is preparing
at this moment several more publications to be submitted in the near future.
After a first pass over
all data taken ($15\cdot10^9$ interactions, $10^9$ events on tape) in
1998/9, from which all of the publications to date
were obtained, we finished a second
pass with improved reconstruction software in 2001.  The analysis of
these data is just starting and many more results are expected.

\section*{Acknowledgments}
The authors are greatful for stimulating discussions with
the SELEX collaboration members.  They wish to thank the
technical staff of their Institutes for their support.
This work was supported by the US Department of Energy under
contract NO.\ DE-AC02-76CHO3000,
the Russian Ministry of Science and Technology,
and Consejo Nacional de Ciencia y Tecnolog\'{\i}a (Mexico).


\begin{thebibliography}{99}

\bibitem{e781} SELEX Collaboration:
Carnegie-Mellon University,
Fermi National Accelerator Laboratory,
University of Iowa,
University of Rochester,
University of Hawaii,
University of Michigan--Flint,
Ball State,
Petersburg Nuclear Physics Institute,
ITEP (Moscow),
IHEP (Protvino),
Moscow State University,
University of S\~ao Paulo,
Centro Brasileiro de Pesquisas Fisicas (Rio de Janeiro),
Universidade Federal da Paraiba,
IHEP (Beijing),
University of Bristol,
Tel Aviv University,
Max-Planck-Institut f\"ur Kernphysik (Heidelberg),
University of Trieste,
University of Rome ``La Sapienza'', INFN,
Universidad Aut\'onoma de San Luis Potos\'{\i},
Bogazici University. \\
{\tt http://fn781a.fnal.gov}

\bibitem{SELEX}  SELEX Collaboration, J.S. Russ {\sl et al.}, 
             in: {\it Proceedings of the 29th International Conference
                 on High Energy Physics}, 1998, 
                 edited by A. Astbury {\sl et al.} 
                 (World Scientific, Singapore, 1998), Vol.\ II, p.\ 1259.
\ifpreprint Preprint hep-ex/9812031.\fi

\bibitem{proto}
M.P.~Maia et al.,
Nucl.\ Instr.\ and Meth. {\bf A326} (1993) 496.

\bibitem{sphinx}
V.A.~Dorofeev et al., Physics of Atomic Nuclei {\bf 57} (1994) 227.\\
A.~Kozhevnikov et al., Nucl.\ Instr.\ and Meth.\ {\bf A433} (1999) 164-167.


\bibitem{elba} J.~Engelfried et al.,
Nucl.\ Instr.\ and Meth.\ {\bf A409} (1998) 439.
\ifpreprint
Preprint FERMILAB-Conf-97/210-E.
\fi

\bibitem{bignim} J.~Engelfried et al.,
Nucl.\ Instr.\ and Meth.\,{\bf A431} (1999) 53-69.
\ifpreprint
Preprint hep-ex/9811001, FERMILAB-Pub-98/299-E.
\fi

\bibitem{israel} J.~Engelfried et al.,
Nucl.\ Instr.\ and Meth. {\bf A433} (1999) 149-152.
\ifpreprint
Preprint FERMILAB-Conf-98/399-E.
\fi


\bibitem{Ronchi} L.~Stutte, J.~Engelfried and J.~Kilmer,
 Nucl.\ Instr.\ and Meth. {\bf A369} (1996) 69.
\ifpreprint
Preprint FERMILAB-PUB-95-138-E.
\fi

\bibitem{CROS} A.~Atamanchuk, V.~Golovstov, L.~Uvarov,
'CROS -- Coordinate
Readout System for multiwire 
proportional  chambers', LNPI Research Report
1988-1989, Gatchina, 1990.

\bibitem{desrich} J.~S\'eguinot and T.~Ypsilantis,
Nucl.\ Instr.\ and Meth. {\bf 142} (1977) 377.

\bibitem{likeli} U.~M\"uller et al, 
Nucl.\ Instr.\ and Meth. {\bf A343} (1994) 279.

\bibitem{uwe}
U.~Dersch et al., Nucl.\ Phys.\ {\bf B579} (2000) 277-312.
\ifpreprint
Preprint Fermilab-Pub-99/325-E, hep-ex/9910052.
\fi

\bibitem{a2}
V.~Molchanov et al., Physics Letters {\bf B521} (2001) 171-180.
\ifpreprint
Preprint Fermilab-Pub-01/256-E, IHEP 2001-34, hep-ex/0109016
\fi

\bibitem{sigmaradius}
I.~Eschrich et al., Physics Letters {\bf B522} (2001) 233-239.
\ifpreprint
Preprint Fermilab-Pub-01/118-E, hep-ex/0106053.
\fi

\bibitem{aharon}
A.~Ocherashvili et al., accepted for publication in Phys.\ Rev.\ C,
hep-ex/0109003.

\bibitem{lclife}
A.~Kushnirenko et al.,
Phys.\ Rev.\ Letters {\bf 86} (2001) 5243-5246.
\ifpreprint
Preprint hep-ex/0010014.
\fi

\bibitem{dslife}
M.~Iori et al.,
Physics Letters {\bf B523} (2001) 22-28.
\ifpreprint
Preprint Fermilab-Pub-01/086-E, hep-ex/0106005.
\fi

\bibitem{lcprod}
F.~Garcia et al.,
Physics Letters {\bf B528} (2002), 49-57.
\ifpreprint
Preprint Fermilab-Pub-01/258-E, hep-ex/0109017.
\fi

\bibitem{Zacatecas}
J.~Russ, in: Particles and Fields, Eighth Mexican Workshop, Zacatecas,
Mexico, 2001. AIP Conference Proceedings 623 (Ed.\ D\'{\i}az-Cruz, Engelfried,
Kirchbach, Mondrag\'on), pp.~163-172.

\bibitem{cabsupxi}
S.Y.Jun et al.,
Phys.\ Rev.\ Letters {\bf 84} (2000) 1857-1861.
\ifpreprint
Preprint Fermilab-Pub-99/217-E, hep-ex/9907062.
\fi

\bibitem{ccu}
M.~Mattson et al., Phys.\ Rev.\ Letters {\bf 89} (2002) 112001.
\ifpreprint
Preprint hep-ex/0208014.
\fi

\end{thebibliography}
\end{document}